\documentclass[12pt,preprint]{aastex}
\usepackage{amsmath}
\usepackage{graphicx}
\usepackage{color}

\newcommand{\km}{\,{\rm km}} \newcommand{\cm}{\,{\rm cm}}
\newcommand{\erg}{\,{\rm erg}} \newcommand{\yr}{\,{\rm yr}}
\newcommand{\ps}{\,{\rm s}^{-1}}
\newcommand{\kpc}{\,{\rm kpc}}
\newcommand{\parsec}{\,{\rm pc}}

\newcommand{\snr}{CTB\,87}
\newcommand{\NHH}{N({\rm H}_2)}
\newcommand{\twCO}{$^{12}$CO}
\newcommand{\thCO}{$^{13}$CO}
\newcommand{\CeiO}{C$^{18}$O}
\newcommand{\otz}{$J$=\,1--0}
\newcommand{\VLSR}{V_{\rm LSR}}
\newcommand{\du}{d_{6.1}}

\newcommand{\E}[1]{\times 10^{#1}}

\begin{document}
\begin{tiny}

\end{tiny}
\title{An Investigation of the Interstellar Environment of Supernova Remnant \snr}
\author{Qian-Cheng Liu$^1$; Yang Chen$^{1,2,8}$; Bing-Qiu Chen$^3$; 
Ping Zhou$^{1,4}$; Xiao-Tao Wang$^{1,5}$; Yang Su$^{6,7}$}
\affil{$^{1}$Department of Astronomy, Nanjing University, 
163 Xianlin Avenue, Nanjing 210023, China}
\affil{$^{2}$Key Laboratory of Modern Astronomy and Astrophysics,
Nanjing University, Ministry of Education, Nanjing 210093, China}
\affil{$^{3}$South-Western Institute for Astronomy Research, Yunnan University, Kunming,
Yunnan 650091, P.R. China}
\affil{$^{4}$Anton Pannekoek Institute, University of Amsterdam,
PO Box 94249, 1090 GE Amsterdam, The Netherlands}
\affil{$^{5}$Department of Physics and Astronomy, University of 
Alabama, 308 Gallalee Hall, Tuscaloosa, AL 35487, USA}
\affil{$^{6}$Purple Mountain Observatory, CAS, 2 West Beijing Road,
Nanjing 210008, China}
\affil{$^{7}$Key Laboratory of Radio Astronomy, Chinese Academy of 
Sciences, Nanjing 210008, China}
\affil{$^{8}$ Corresponding author; ygchen@nju.edu.cn}

\begin{abstract}

We present a new millimeter CO-line observation toward 
supernova remnant (SNR) \snr, which was regarded purely as a
pulsar wind nebula (PWN),
and an optical investigation of a coincident surrounding superbubble.
The CO observation shows that
the SNR delineated by the radio emission is projectively
covered by a molecular cloud (MC) complex
at $\VLSR=-60$ to $-54\km\ps$.
Both the symmetric axis of the radio emission and the
trailing X-ray PWN appear projectively to be along a gap
between two molecular gas patches at $-58$ to $-57\km\ps$.
Asymmetric broad profiles of \twCO\ lines peaked at $-58\km\ps$
are found at
the eastern and southwestern edges of
the radio emission.
%This very probably represents a kinematic evidence of the
%SNR-MC interaction.
This represents a kinematic signature consistent with an
SNR-MC interaction.
We also find that a superbubble, $\sim37'$ in radius,
appears to surround the SNR
from HI 21cm ($\VLSR\sim-61$ to $-68\km\ps$), \emph{WISE} mid-IR,
and optical extinction data.
We build a multi-band photometric stellar sample of
stars within the superbubble region and find 82 OB star candidates.
The likely peak
distance in the stars' distribution seems consistent
with the distance previously suggested for \snr.
We suggest the arc-like radio emission is mainly a relic of
the part of blastwave
that propagates into the MC complex and is now in a radiative stage
while the other part of blastwave has been expanding into
the low-density region in the superbubble.
This scenario naturally explains the lack
of the X-ray emission related to the ejecta and blastwave.
The SNR-MC interaction also favors a hadronic contribution
to the $\gamma$-ray emission from the \snr\ region.

\end{abstract}


\section{Introduction} 

The progenitors of core-collapse (CC) supernovae (SNe) are massive
stars, which are expected to be born mostly in OB associations,
and their environments are altered by energy feedback processes
dominated by energetic stellar winds and SN explosions
as well as strong ionizing radiation.
Such processes can produce superbubbles filled with 
low-density material and significantly affect the evolution of 
supernova remnants (SNRs) therein.
On the other hand, remnants of CC SNe,
are often close (a few pc)
to molecular clouds (MCs), the birthplace of the
progenitor stars that end their evolution in short lifetimes.
So far about 70 SNRs are confirmed or suggested
to be associated with MCs \citep{2010ApJ...712.1147J,2014IAUS..296..170C}
among the known $\sim300$ SNRs in the Milky Way 
\citep{2012AdSpR..49.1313F, 2014BASI...42...47G}.
Six types of observational evidence for SNR-MC interaction are summarized,
including the 1720 MHz OH maser, molecular line broadening,
morphological agreement of molecular features with SNR features, etc.\
\citep{2010ApJ...712.1147J, 2014IAUS..296..170C}.
In this paper, we investigate a perplexing SNR, \snr,
that may interact with an adjacent MC but may be located near 
the inner edge of a superbubble.

\snr\ has been classified as a filled-center type SNR,
with a radio size of about $8'\times6'$, centered at
R.A.$= 20^{\rm h}16^{\rm m}02^{\rm s}$, decl.$= 37^{\circ}12'00''$.
It was first cataloged in a radio survey at 960\,MHz of the Galactic plane
conducted by the Owens Valley Radio Observatory \citep{1960PASP...72..331W}.
The distance to it was suggested to be $12\kpc$
according to HI absorption measurements
\citep{1989MNRAS.237..555G,1997A&A...317..212W}.
Based on the extinction-distance relation \citep{2002ASPC..276..123F},
a new distance of $6.1\pm0.9\kpc$ was established,
which implies that \snr\ is located in the Perseus spiral arm
\citep{2003ApJ...588..852K}.
In X-rays, \snr\ is centrally brightened with a size of about $5'$
as observed by the \emph{Einstein} satellite observatory
\citep{1980ApJ...241L..19W}. 
A detailed \emph{Chandra} ACIS analysis shows that the pulsar
wind nebula (PWN) harbours
a point source, which may putatively be a pulsar,
$100''$ southeast away from the radio emission peak
\citep{2013ApJ...774...33M}.
In $\gamma$-rays, the TeV point source VER J2016+371,
spatially coincident with \snr, is resolved
by the VERITAS telescope system \citep{2014ApJ...788...78A}.
The GeV \emph{Fermi}-LAT source 3FGL J2015.6+3709
is spatially close to VER J2016+371,
with its origin still under debate \citep{2012ApJ...746..159K,2015ApJS..218...23A,
2016ApJS..224....8A,2016MNRAS.460.3563S}.
Recently, the multiwavelength spectrum obtained from the \snr\ region
has been interpreted to be comprised of contributions from a Maxwellian
distribution of electrons and a broken power-law distribution of electrons
\citep{2016MNRAS.460.3563S}.

\snr\ has been proposed to be associated with molecular gas at a systemic
local standard of rest (LSR) velocity $\VLSR$ of
about $-58\km\ps$ \citep{1986ApJ...309..804H, 1994AJ....108..634C,2003ApJ...588..852K},
while it has also been argued to lie at the inner boundary 
of an expanding HI superbubble at a $\VLSR$ of
about $-70\km\ps$ \citep{1997A&A...317..212W}.
The molecular gas was resolved into 
several subclumps, with \snr\ seemingly located between
two of them \citep{1994AJ....108..634C}.
However, no kinematic evidence has yet been found for the SNR-MC interaction.
Also, no 1720\,MHz OH maser, a reliable signpost of interaction between 
SNR and MC, has been detected for \snr\ \citep{1996AJ....111.1651F}.
About 30\% of the SNRs that are confirmed to be in physical contact with
MCs do not show OH masers \citep{2010ApJ...712.1147J}, most probably
because the shocked molecular gas does not satisfy the appropriate
physical conditions, namely density of order $\sim 10^5 {\rm cm}^{-3}$
and temperature in the range of $50$--$125$\,K \citep{1999ApJ...511..235L}.

In previous work, \cite{1994AJ....108..634C} mapped
the \snr\ region in CO lines with a half-power beamwidth of $2.7'$
and a grid spacing of $3'$,
and \cite{2003ApJ...588..852K} presented a mapping in CO~(\otz)
with a beamwidth of $2'$ and a grid spacing of $1'$,
as well as a partial mapping in CO~($J$=\,3--2).
In this paper, we present a new CO-line observation toward
the \snr\ region and an optical investigation of the
coincident superbubble, aiming to explore the environment of \snr\
and the interaction of the SNR with the adjacent MC.
We first describe the CO observations and data reduction process 
in \S\ref{data}; we analyze the CO-line data in detail in \S\ref{result} and 
discuss the main result in \S\ref{discussion}. 
Finally, we present our summary in \S\ref{summarize}.

\section{Observations and data reduction} \label{data}

Our observations of millimeter emission toward SNR~\snr\ were made
simultaneously in \twCO\ (\otz), \thCO\ (\otz), and \CeiO ($J=$1--0)
lines during 2013 May 11--18 using the 13.7m millimeter-wavelength telescope
of the Purple Mountain Observatory at Delingha.
We used the new $3\times3$ pixel
Superconducting Spectro-scopic Array Receiver as the front end,
which is constructed with Superconductor-Insulator-Superconductor mixers
using the sideband separating scheme
\citep{2011AcASn..52..152Z, 2012ITTST...2..593S}.
We did on-the-fly mapping toward a $34'\times60'$ field that covers
the full angular extent of SNR \snr\
with a grid spacing of $30''$.
The half-power beamwidth of the telescope is about $52''$
and the pointing accuracy of the telescope is better than $4''$.
The typical system temperature is about 220K for 115.2GHz
and 130K for 110.2GHz.
The bandwidth of the spectrometer is 1\,GHz,
with 16,384 channels and a spectral resolution of 61\,KHz.
Thus the velocity resolution is $0.16\km\ps$ for \twCO\
and $0.17\km\ps$ for \thCO\ and \CeiO.
All the CO-line data were reduced using the GILDAS/CLASS package
developed by the IRAM observatory\footnote{http://www.iram.fr/IRAMFR/GILDAS}. 

For a multiwavelength investigation of the environment of this SNR,
we also used \emph{Chandra} X-ray (ObsID: 11092, PI: Safi-Harb;
which we reduced with CIAO ver.\ 4.7),
and \emph{WISE} 22 $\mu$m mid-infrared (IR) (\emph{WISE} 
Science Data Center, IPAC, Caltech) data.
The HI line emission data were obtained from the 
Canadian Galactic Plane Survey
\citep[CGPS;][]{2003AJ....125.3145T},
and the 1.4\,GHz radio continuum emission data
were from the NRAO VLA Sky Survey \citep[NVSS;][]{1998AJ....115.1693C}.
Optical and near-IR data from the 
Panoramic Survey Telescope and
Rapid Response System \citep[Pan-Starrs;][]
{2002SPIE.4836..154K},
the Isaac Newton Telescope (INT) Photometric ${\rm H}\alpha$
Survey of the Northern Galactic Plane
\citep[IPHAS;][]{2005MNRAS.362..753D},
and the Two Micron All Sky Survey
\citep[2MASS;][]{2006AJ....131.1163S}
were used to investigate the OB star candidates within the
coincident superbubble.

\section{Results} \label{result}

\subsection{Spatial distribution of the ambient clouds}
\label{spatial}
\begin{figure}
\centering
\includegraphics[scale=.7]{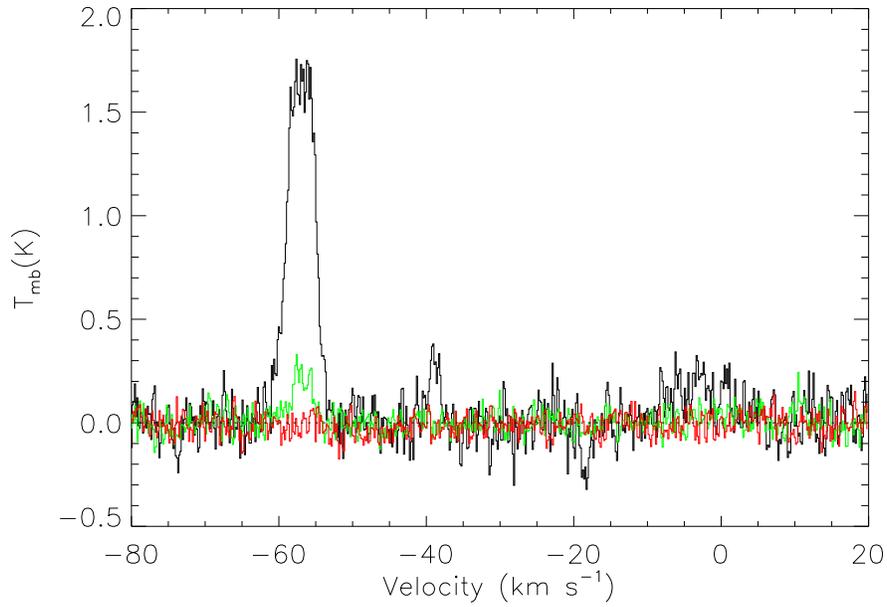}
\caption{Average CO spectra from a $6.5'\times6.5'$ region
(centered at R.A.=$20^{\rm h}16^{\rm m}06^{\rm s}.7$, 
decl.=$37^{\circ}12'32''$)
covering the pulsar wind nebula,
in the velocity range of 
$-80 \km \ps$-- $20 \km \ps$. The black line is for \twCO\ (\otz),
the green line for \thCO\ (\otz), and the red line for \CeiO (\otz).
}
\label{overallspec}
\end{figure}

Figure~\ref{overallspec} shows the averaged CO
spectra from a region covering SNR \snr.
There are two prominent \twCO~(\otz) emission peaks, 
at around $-58 \km\ps$ and $-38 \km\ps$.
The \thCO~(\otz) emission is only prominent at $\sim-58\km\ps$,
and the \CeiO~(\otz) emission is not significant across the velocity range.

\begin{figure}
\centering
\includegraphics[scale=.8]{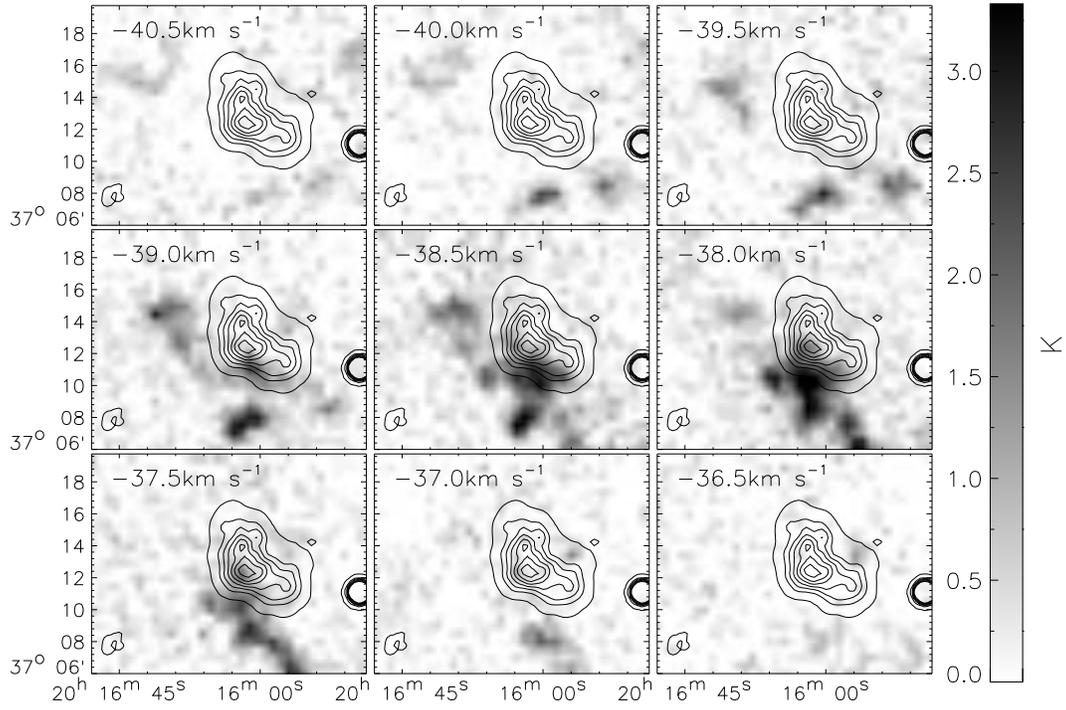}
\caption{\twCO\ intensity maps integrated each $0.5 \km \ps$
in the velocity range of $-40.5$ to $-36.5 \km \ps$,
overlaid by NVSS 1.4 GHz radio continuum emission in gray contours
with levels of 4, 45, 86, 127, 168, 209, and 250 mJy beam$^{-1}$.
The lowest contour level is larger than the 
5$\sigma$ value of the background.
\label{channelmap1}}
\end{figure}

\begin{figure}
\centering
\includegraphics[scale=.8]{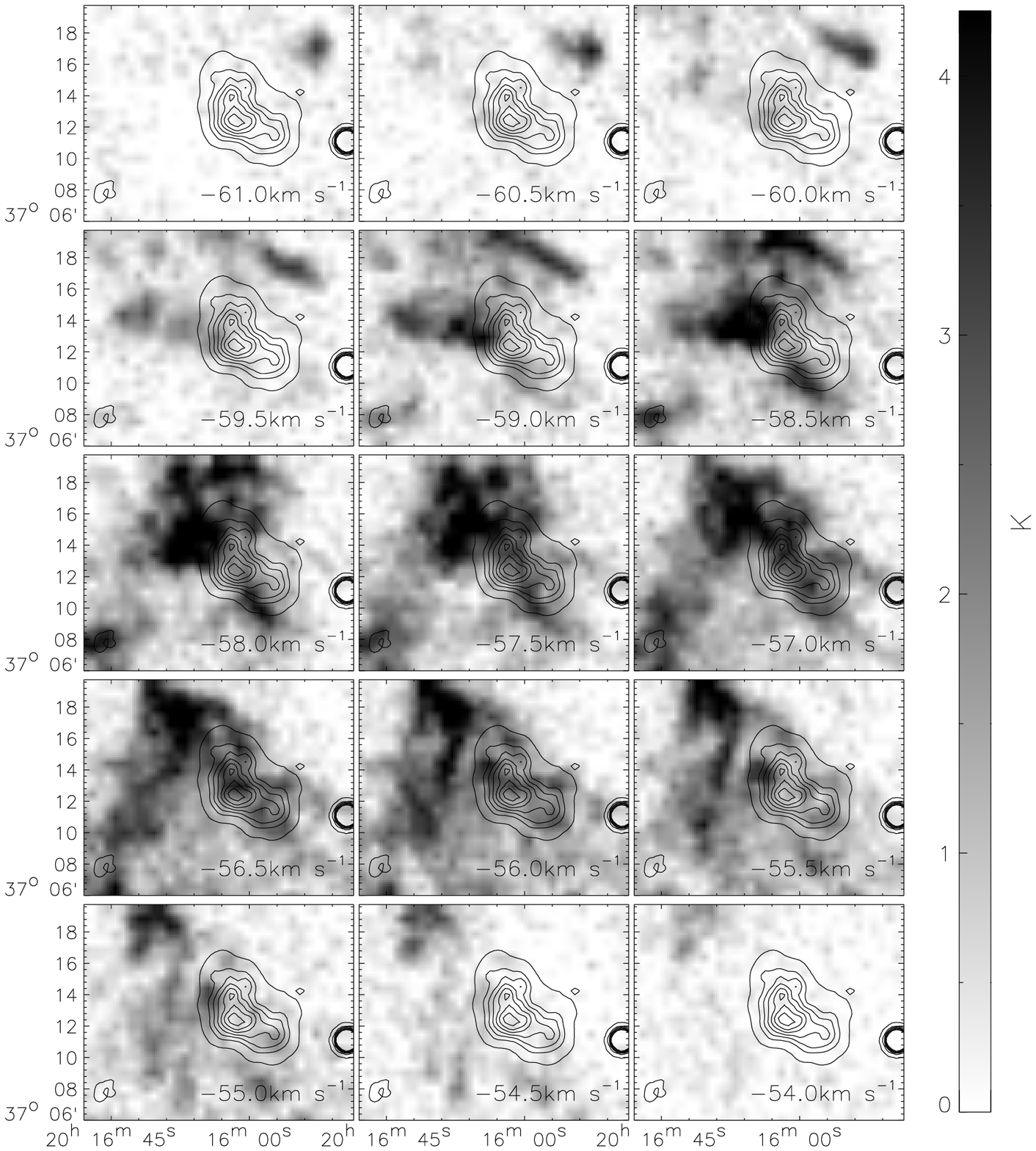}
\caption{\twCO\ intensity maps integrated each $0.5 \km \ps$
in the velocity range $-61$ to $-54 \km \ps$.
The contours are the same as those in 
Figure \ref{channelmap1}.
\label{channelmap2}}
\end{figure}

We made \twCO\ emission channel maps around the two 
peaks with velocity interval $0.5\km\ps$
(see Figure~\ref{channelmap1} and Figure~\ref{channelmap2})
to examine the spatial distribution of the two molecular components.
The $\sim-38\km\ps$ \twCO\ component seems to overlap the SNR
at its southeastern ``apex"
by projection in interval $-39$ to $-37.5\km\ps$, but there is no
systematic morphological feature correspondence between the molecular gas
and the SNR (Figure~\ref{channelmap1}).

\begin{figure}
\centering
\includegraphics[scale=.8]{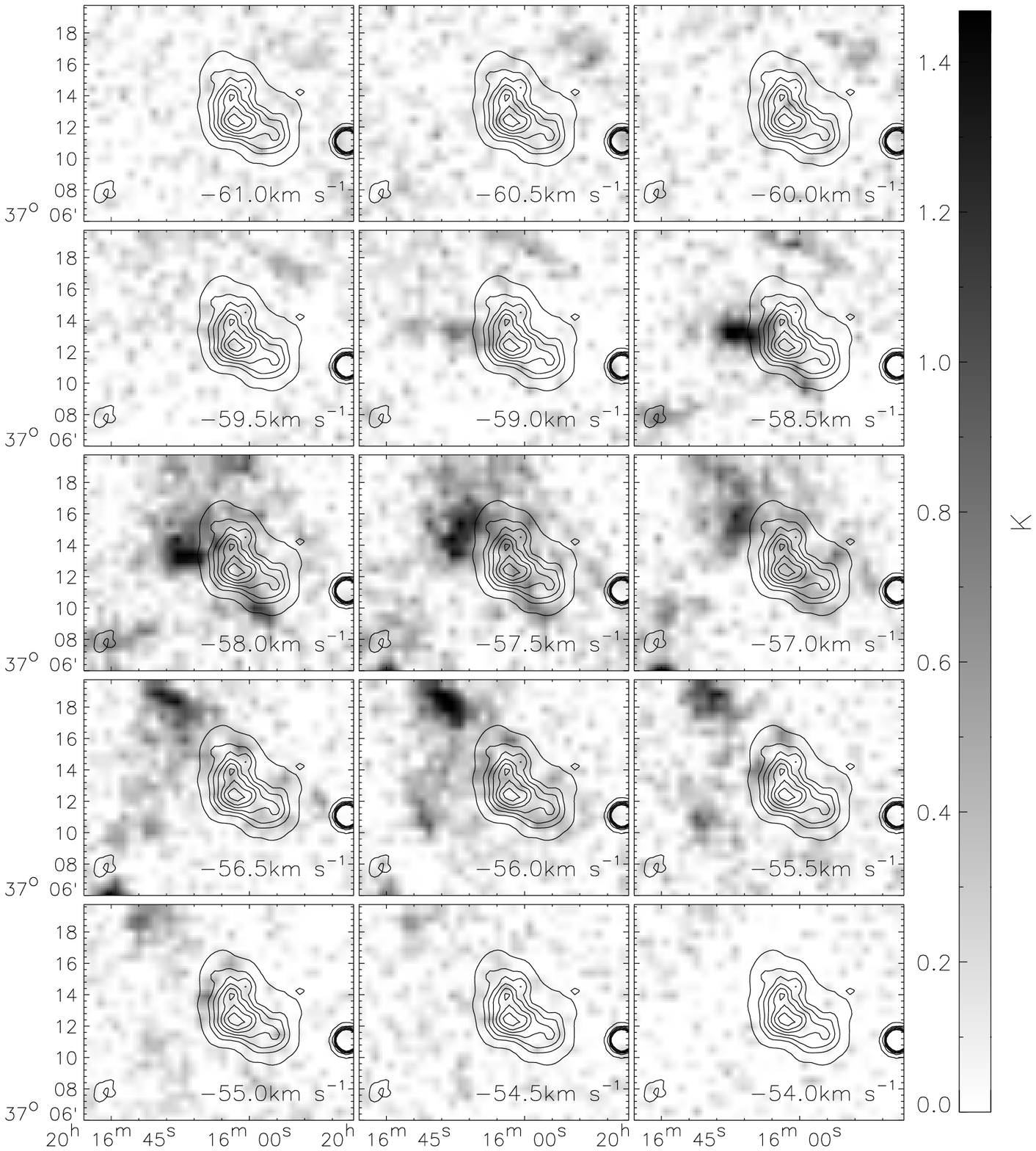}
\caption{\thCO\ intensity maps integrated each $0.5 \km \ps$
in the velocity range of $-61$ to $-54 \km \ps$.
The contours are the same as those in 
Figure \ref{channelmap1}.   \label{channelmapl}}
\end{figure}

\begin{figure}
\centering
\includegraphics[scale=.5]{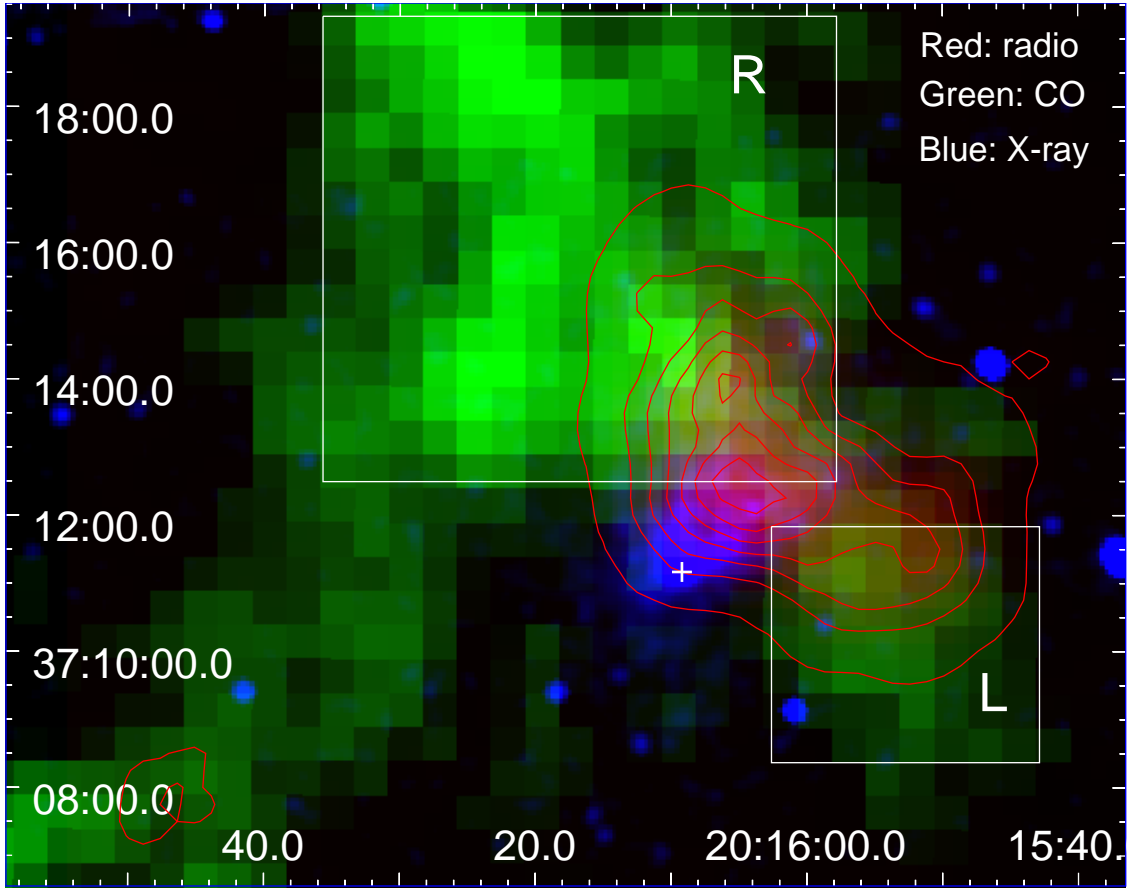}
\caption{Multiwave map of the SNR \snr: NVSS 1.4 GHz radio 
continuum emission in red, \twCO\ (\otz) intensity map in 
velocity interval $-60$ to $-54 \km \ps$ in green, 
\emph{Chandra} X-ray image in energy band 0.5--7 keV in blue. 
We have also overlaid the contours of the NVSS 1.4 GHz 
radio continuum emission with levels the same as in
Figure \ref{channelmap1}. 
The white cross indicates the point source
CXOU\,J201609.2+371110 reported in
\cite{2013ApJ...774...33M}.
\label{overall}}

\end{figure}

The $\sim-58\km\ps$ component appears to spread over a large area
covering the SNR, and a few LSR-velocity dependent 
structures are noteworthy.
(1)
In velocity interval $-55.5$ to $-54.5\km\ps$,
there is a linear structure
%clearcut wall or shell like structure
of \twCO~(\otz) emission
along the eastern edge of the SNR
(Figure~\ref{channelmap2}),
somewhat similar to that described in \cite{2003ApJ...588..852K}
(in passing, note that a Galactic coordinate system was used there
while an equatorial coordinate system is used here, and 
the shell-like structure shown there is not complete due to
the limited field of view).
The structure can also be discerned in the \thCO\ (\otz) channel map
at $-55\km\ps$ (Figure~\ref{channelmapl}).
(2) In velocity interval $-56.5$ to $-55.5\km\ps$, a bar-like molecular
structure appears to pass through the SNR along the northeast-southwest
orientation and, in particular, through the brightness peak of the
radio emission.
(3) In interval $-58$ to $-57 \km \ps$, both \twCO\ and \thCO\
emissions show two patches of molecular material in the eastern
and southwestern edges of the SNR,
similar to the sub-clumps described in \cite{1994AJ....108..634C}.
The integrated \twCO~(\otz) intensity map in the main velocity range of
the $\sim-58\km\ps$ component, $-60$ to $-54 \km \ps$,
 is presented in Figure~\ref{overall} (in green).
It is overlaid with \emph{Chandra} 0.5--7\,keV X-ray image of \snr\
(in blue) and the radio continuum image (in red).
The X-ray emitting part of the PWN, with a southeast-northwest
oriented elongation, appears to be located in a gap
of molecular gas, where CO emission is weak,
between the two patches at $-58$ to $-57\km\ps$.

\subsection{Kinematic evidence for SNR-MC interaction ?}
\label{kinematic}
\begin{figure}
\centering
\includegraphics[scale=1.]{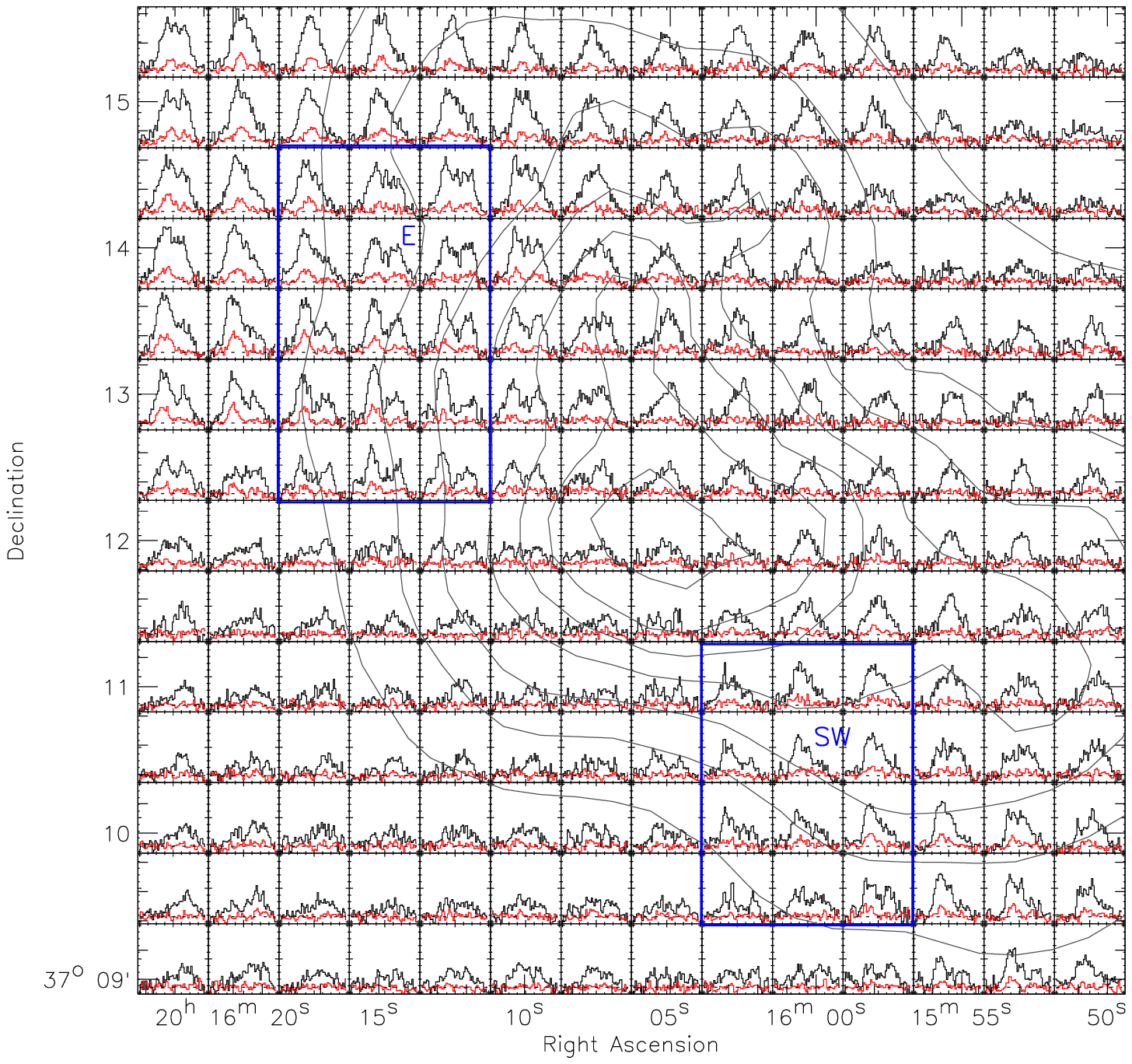}
\caption{Grid of \twCO\ (\otz) and \thCO\ (\otz) spectra 
in the velocity range of $-61$ to $-53 \km \ps$.
Black lines denote \twCO\ spectra, red lines denote \thCO, 
and dashed lines denote the 0\,K main-beam temperature.
The size of each pixel is $30''\times30''$. 
The radio contours are the same as those in Figure \ref{channelmap1}. 
Two regions delineated by blue rectangles (labelled as ``E'' and ``SW'') 
are used to extract CO spectra,
in which redward broadened wings are shown for the \twCO\ emission
at systemic velocity $\sim-58 \km \ps$
(see Figure~\ref{region}). \label{linegrid60}}
\end{figure}

\begin{figure}
\centering
\includegraphics[scale=1.]{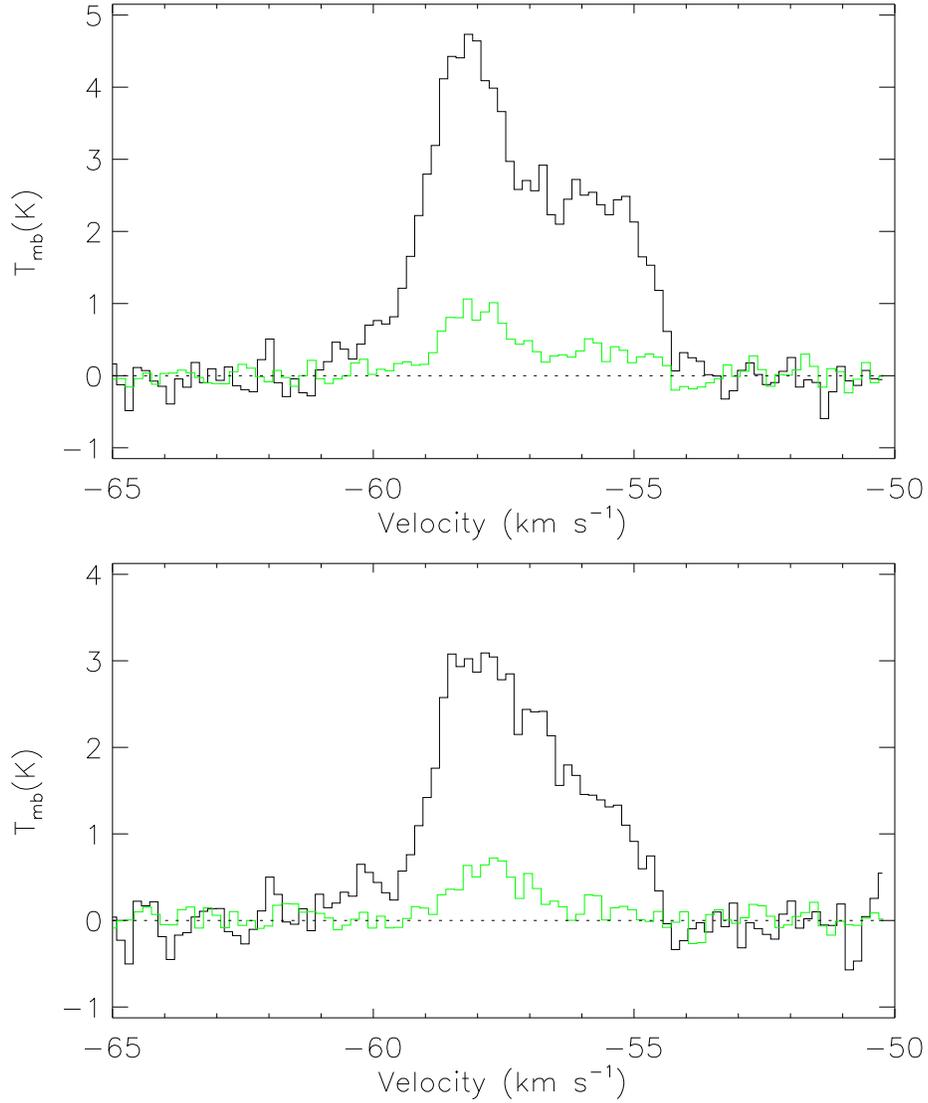}
\caption{Averaged spectra of regions ``E" (top panel) and ``SW"
(bottom panel) in the velocity range $-65$--$-50\km\ps$. 
The two regions are defined in Figure~\ref{linegrid60}.
The black line denotes the \twCO\ spectra, the green line 
represents \thCO, and the dotted line 
represents the 0\,K main-beam temperature. \label{region}}
\end{figure}

We have inspected the \twCO~(\otz) line profiles of 
the two molecular components toward \snr.
For the $\sim-38\km\ps$ component, we do not find any
asymmetric broad profiles of the \twCO\ line in the 
grid of CO spectra.
(This component, as mentioned in \S\ref{spatial}, has neither
morphological feature corresponding to the SNR).
%Little notable sign of broad line is found in the
%$\sim-38\km\ps$ component 
%in the field of view.
For the other component, around $-58\km\ps$,
we show a grid of \twCO\ (\otz) and \thCO\ (\otz) spectra
in the velocity range of $-61$ to $-53 \km \ps$ (Figure~\ref{linegrid60})
and find 
asymmetric broad profiles of
the \twCO\ line in the eastern and southwestern edges.
Figure~\ref{region} shows the average line profiles of some of the pixels
at the edges of the SNR (regions ``E"
and ``SW'' marked in Figure~\ref{linegrid60}).
There is a secondary peak at $-56$ to $-55\km\ps$ in region ``E"
(also seen in most of the pixels therein) besides the main peak
at $-58\km\ps$,
with little \thCO\ counterpart ($\la3\sigma$).
Although the broad red (right) wing may include a contribution
from real line broadening due to shock disturbation,
we cannot exclude contamination by line-of-sight emission
in a wide region.
We note that there is similar secondary peak on the east of the radio boundary
and there is non-negligible intensity at $-56$ to $-55\km\ps$ on the northeast
of the boundary.
However,
in the three bottom pixels of region ``E", there are unique plateaus
in the blue (left) wings ($\sim-61$ to $-59\km\ps$),
which are also clearly ($>4\sigma$) reflected in the average profile for
region ``E" (top panel, Figure~\ref{region}).
This plateau-like spectral feature is somewhat similar to that detected
in the \twCO\ spectra in the western edge of SNR~3C397 \citep{2010ApJ...712.1147J}.
In the spectra of the pixels in the southwestern edge,
as typified by region ``SW'', the line profiles are characterized
by strong peaks at $-58\km\ps$ for both \twCO\ and \thCO\ 
and a broad red wing of \twCO\ extending to $\sim-54\km\ps$\
without a significant \thCO\ counterpart
(bottom panel, Figure~\ref{region}).
The surrounding pixels essentially display different shapes of line profiles,
and the emission outside the southern edge becomes weak.
Since \thCO\ emission, which is usually optically thin
(like that with $\tau$(\thCO)$\ll1$ given in
Tables~\ref{parameter} and \ref{parameter2}),
is yielded in quiescent, intrinsically high-column-density
molecular gas,
a broad \twCO-line wing without a \thCO\ counterpart
is very likely to represent a perturbed gas deviating from the systemic
LSR velocity.
Therefore,
the asymmetric \twCO\ wings from the edges of \snr\ 
(especially the blue wings in region ``E"
and the red wings in region ``SW")
very likely result from Doppler broadening
of the $-58\km\ps$ line
and thus might provide a kinematic evidence for interaction
between the $\sim-58\km\ps$ MC complex and the SNR.

We fit the CO emissions with Gaussian lines for the $\sim-58 \km\ps$
molecular gas in regions ``R" and ``L"
(as defined in Figure~\ref{overall}),
and the derived parameters,
molecular column density $N(\mbox{H}_2)$, excitation temperature
$T_{\rm ex}$, and optical depth of \thCO\,(\otz) $\tau$(\thCO)
are summarized in Table~\ref{parameter}
and Table~\ref{parameter2}. 
The distance to the MC/SNR is taken to be 6.1 kpc, as suggested 
by \cite{2003ApJ...588..852K}. 
The column density of H$_2$ and the mass of the molecular gas
are estimated using two methods.
In the first method, 
the conversion relation for the molecular
column densities,
$N$(H$_2$)\,$\approx 7\times 10^5 N$~(\thCO) \citep{1982ApJ...262..590F},
is used under the assumption of local thermodynamic equilibrium
for the molecular gas and optically thick condition for
the \twCO\ (\otz) line.
In the second method, 
a value of the CO-to-H$_2$ mass conversion factor,
$N$(H$_2$)/$W$(\twCO) (known as the ``X-factor"), 
$1.8\times 10^{20}$\,cm$^{-2}$\,K$^{-1}$\,km$^{-1}$\,s
\citep{2001ApJ...547..792D} is adopted.

\subsection{The superbubble at $\VLSR\sim-64\km\ps$}
\label{superbubble}
\begin{figure}
\centering
\includegraphics[scale=.5]{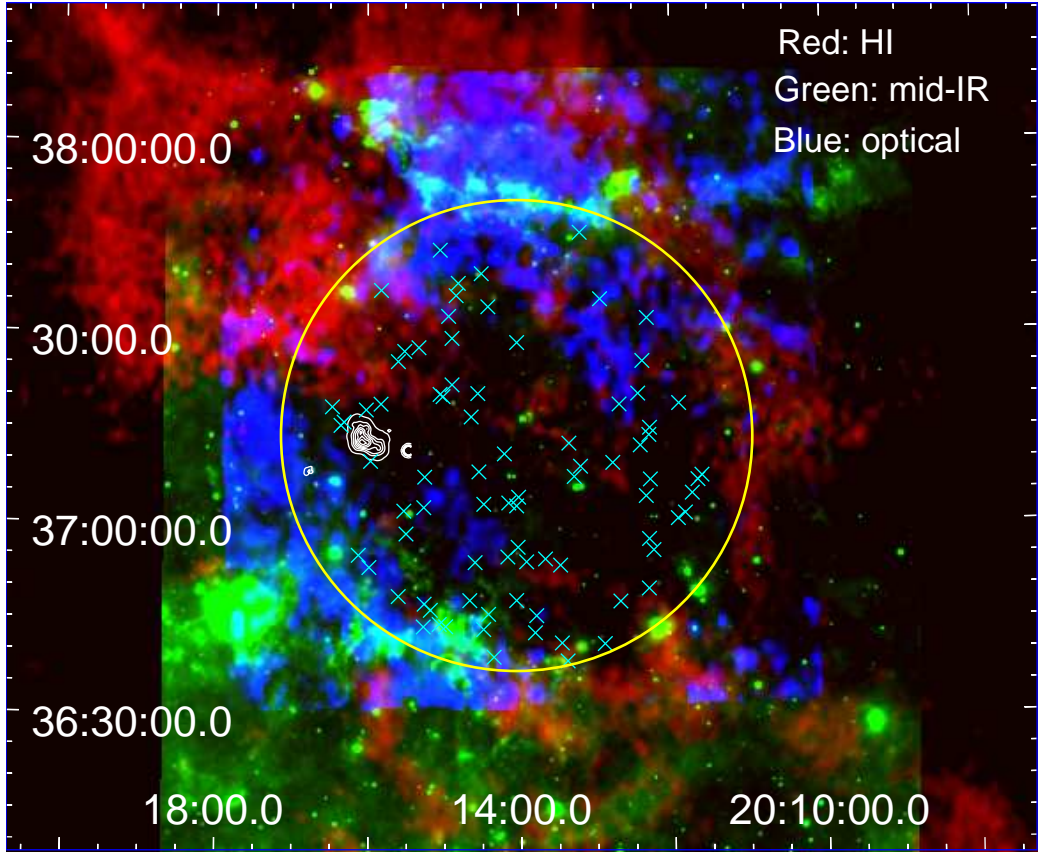}
\caption{Multiwave band morphology of the superbubble 
in the direction of SNR \snr;
HI emission around $\VLSR=-64 \km\ps$ in red,
\emph{WISE} 22.194 $\mu$m mid-IR image in green, 
optical extinction map in blue. 
The radio contours in white are the same 
as those in Figure~\ref{channelmap1}. 
The large yellow circle outlines the superbubble region,
and the cyan crosses mark the project positions of 
the OB star candidates.
\label{bubble}
}
\end{figure}

By projection, SNR \snr\ is located within a superbubble, centered at
R.A.=$20^{\rm h}14^{\rm m}03^{\rm s}$, decl.=$37^{\circ}13'27''$,
with a circular boundary (brightened in the south and north)
of angular radius $\sim37'$, and nears its eastern edge
(shown in Figure~\ref{bubble}). 
We investigate the \emph{WISE} mid-IR observation toward
SNR~\snr\ and find the superbubble is 
bright at both 12 and $22\mu$m.
The HI emission in this sky region also shows a cavity at
the LSR velocity around $-64 \km\ps$ ($-61$\,to\,$-68\km\ps$),
which happens to be spatially coincident with the mid-IR superbubble
and to have the same angular size. 
Furthermore, such a bubble-like structure has a counterpart
in the optical extinction map (also see Figure~\ref{bubble}).

As superbubbles are usually the products of energy and material feedback
from massive OB stars into the interstellar medium,
we searched for possible OB stars 
in the direction toward this superbubble. 

We built a multi-band photometric stellar sample of 
over 0.18~million 
stars within the projected region of the superbubble
(i.e., the circle shown in Figure~\ref{bubble})
by cross-matching the photometric cataloges of the 
Pan-Starrs, IPHAS, and 2MASS.
We used a spectral energy distribution (SED) fitting algorithm, 
similar to those employed by \cite{2014MNRAS.443.1192C} 
and \citet{2015MNRAS.450.3855M}, to obtain the 
effective temperature $T_{\rm eff}$, 
optical extinction $A_V$, and distance modulus $\mu$ 
of the individual stars.
We compared the observed SED of each star 
to the stellar models from the Padova isochrone data base 
\citep[CMD v3.0;][]{2012MNRAS.427..127B,2017ApJ...835...77M}.
We considered only the main-sequence 
(surface gravity $\log g> 3.8$\,dex),
and 
Galactic thin disk metallicity ([Fe/H] = $-0.2$\,dex)
models. The observed magnitudes of a star can be modeled by 
\begin{equation}
  m^{\rm obs}_i=m^{\rm mod}_i+A_i+\mu,
\end{equation}
where $m^{\rm obs}_i$ and $m^{\rm mod}_i$ are the observed 
and Padova magnitudes in the $i$th band 
($i$=1, 2, ..., 11 corresponding to Pan-Starrs 
$g_{\rm P},~ r_{\rm P},~i_{\rm P},~z_{\rm P},~y_{\rm P}$, 
IPHAS $r',~i'$, H$\alpha$, and 2MASS 
$J$, $H$, and $K_{\rm S}$ bands, respectively),
$A_i$ is the extinction in the $i$th band, and 
$\mu$ is the distance modulus. 
We adopt the $R_V=3.1$ extinction law 
from \cite{1989ApJ...345..245C} to convert the optical extinction 
$A_V$ to extinction in each individual band $A_i$.
A simple Bayesian scheme based on the Markov Chain Monte Carlo 
sampling is adopted to obtain the parameters 
$\log T_{\rm eff}$, $A_V$, and $\mu$ for the individual stars. 
We adopt the likelihood 
\begin{equation}
  L_r=\Pi  \frac{1}{\sqrt{2\pi \sigma _i}}
  \exp{\frac{-(m^{\rm obs}_i-m^{\rm mod}_i)^2}{2\sigma^2 _i}},
\end{equation}
where $\sigma_i$ is the uncertainty of the $i$th band observed magnitude,
and the priors 
\begin{multline}
P(T_{\rm eff}, A_V, \mu) =   \begin{cases} 1 \quad \text{if} \,\,\begin{cases} 4.3 \,\, \leq \,\, {\rm log}\,(T_{\rm eff}) \,\, \leq \,\, 4.7 \\
0 \,\, \leq \,\, A_V \,\, \leq \,\, 20 \\
0 \,\, \leq \,\, \mu \,\, \leq \,\, 20
 \end{cases}\\
 0 \quad \text{else}.
 \end{cases}
\end{multline}
We adopt the results for the stars 
with high photometric precisions 
($\sigma_i$ $\le$ 0.1\,mag for each band) and high
posterior probabilities ($\log P\ge 5$, 
where 5 is around the median value of
the logarithmic posterior probabilities).
We note that, due to the lack of information in blue optical 
wavelengths (i.e.,  $u$ band), there could be 
degeneracies between the effective temperatures and the 
amount of extinction for the individual stars. 

\begin{figure}
\centering
\includegraphics[scale=1.0]{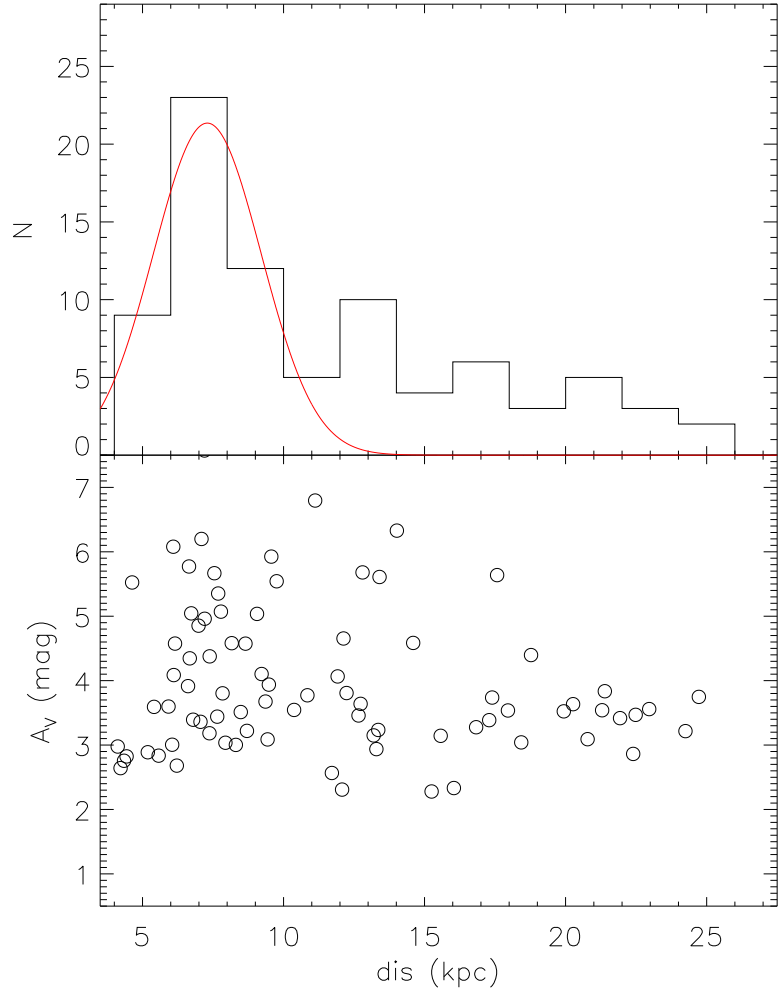}
\caption{
Distribution of distance (top panel) 
and optical extinction (bottom panel) 
for the OB star candidates ($\log P\ge 5$ and $\log T_{\rm eff}\ge 4.3$)
within the projected region of the superbubble 
(i.e., the circle shown in Figure~\ref{bubble}). 
The red curve in the top panel is the Gaussian fitting result
of the peak-like component.
\label{obstar}
}
\end{figure}

As a result,
we found 82 OB star candidates.
Figure~\ref{obstar} shows the distribution of 
distance and optical extinction for all these objects.
%It is high likely that there is a peak
There seems to be a peak
%(at a $\sim 5\sigma$ level)
of the distribution of distance
of these OB candidates at $d$=7.3\,kpc, with a dispersion of 1.9\,kpc,
%Therefore the OB association that creates the superbubble
%is most likely indicated by the peak,
although the selected stars may be an incomplete and contaminated
sample of the candidates owing to the lack of the $u$ band data.
Notably, this distance range covers that of the Perseus spiral arm 
in the direction of \snr\
\citep[][Figure\,8 therein]{2003ApJ...588..852K}.
As some massive stars should be responsible for the superbubble,
it is most likely that they are located in the dispersion range
of the peak.
The distance range of the massive stars
or the superbubble 
also seems to be consistent with the distance to SNR~\snr, 
either $6.1 \pm 0.9$\,kpc, that is derived from the intervening
neutral hydrogen column density \citep{2003ApJ...588..852K}
or $\sim6$--8\,kpc estimated from the column density of the
line-of-sight X-ray absorbing hydrogen atoms and molecules
(see \S~\ref{snrenvironment}).

\section{Discussion} \label{discussion}

\subsection{SNR Environment}\label{snrenvironment}
Our new CO-line observation shows spatial correspondence of
the SNR \snr\ with an MC complex at LSR velocities
around $-58\km\ps$ ($-54$ to $-60\km\ps$).
We have also demonstrated a probable kinematic 
evidence of the SNR shocking against molecular patches
at $\VLSR\sim-58\km\ps$.
Hereafter we parameterize the distance to \snr\ and
the associated MC as $d=6.1\du\kpc$.
We can make a rough estimate of the density of the associated molecular gas.
We adopt angular sizes of the two patches of molecular gas in regions
``R" and ``L" as
$\sim 7.5'\times 7.5'$ and $\sim 3.9'\times 3.5'$
(i.e., $\sim 13.3\du\parsec\times 13.3\du\parsec$ and $\sim 6.9\du\parsec\times 6.2\du\parsec$), respectively,
and assume line-of-sight sizes
13.3\,pc and 6.6\,pc for them.
The estimated gas density, $n(\mbox{H}_2)$, and gas mass, $M$, 
for the two patches are given in Tables~\ref{parameter} and \ref{parameter2},
respectively.

The multiwavelength spectrum from radio to $\gamma$-rays for \snr,
which incorporates the TeV emission of VER\,J2016+371 \citep{2014ApJ...788...78A} and
the GeV emission from 3FGL J2015.6+3709 
\citep{2012ApJ...746..159K,2015ApJS..218...23A,2016ApJS..224....8A},
has been explained using a
leptonic scenario by \cite{2016MNRAS.460.3563S},
with a combination of a broken power law and a Maxwellian 
distribution of electrons.
Nonetheless, it is also mentioned that the observed GeV--TeV data can be explained
by a hadronic scenario as long as there is a dense ambient target medium
with a mean hydrogen atom density $\sim 20\,{\rm cm}^{-3}$ or higher
(depending on detailed hadronic interaction mechanisms)
\citep{2016MNRAS.460.3563S}. 
Actually, it is notable that the centroid of the {\sl VERITAS} 
very high energy $\gamma$-ray
emission \citep{2014ApJ...788...78A} is essentially coincident with 
both the radio emission of the SNR and the $\sim-58\km\ps$ MC complex.
The possible SNR-MC interaction revealed here provides a likely hotbed
for the production of hadronic $\gamma$-rays.
The gas density of the two molecular patches,
a few tens cm$^{-3}$, estimated here, appears to satisfy the hadronic scenario.

\snr\ is very likely to be located within (but near the eastern boundary of)
 the revealed superbubble, 
in view of the following issues.
(1) The LSR velocity of the \twCO\ emission of the associated MC, 
$-54$ to $-60\km\ps$,
is very similar to that of the HI emission for the superbubble,
$-61$ to $-68\km\ps$.
(2) 
The distance to the SNR, derived both from the foreground 
intervening neutral hydrogen column density, $6.1\pm0.9\kpc$
\citep{2003ApJ...588..852K}, 
and from the X-ray-absorbing hydrogen column density,
$6$ to $8\kpc$ (see below),
is consistent with the likely peak distance of the
OB star candidates that are projected inside the superbubble,
7.3$\pm$1.9\,kpc,
which can be regarded as the distance to the superbubble
and the 
OB star candidates inside. 
%responsible OB association.
The hydrogen (including atoms and molecules) column density for the entire diffuse
X-ray emission region of \snr\
is $N_{\rm H}=1.21$--$1.57\E{22}\cm^{-2}$
\citep{2013ApJ...774...33M}.
The optical extinction can then be estimated to be
$A_V=4.8\pm 0.7$ from the empirical relation
$N_{\rm H}=(2.87\pm0.12)\E{21}A_V\cm^{-2}$
\citep{foight16},
which indicates a distance of $\sim6$--8\,kpc to \snr\
using the $A_V$-distance relation toward this SNR given in \cite{2003ApJ...588..852K}.
(3) There is little detection of thermal X-rays 
related to the SN ejecta and blast wave, 
which sets an upper limit $0.2\cm^{-3}$ to the density of the
medium \citep{2013ApJ...774...33M}.
This can be readily explained if the SN exploded material
is blown into the tenuous and hot gas within the superbubble
(in \snr, another part of the gas may have shocked into the MC; see \S\ref{configuration}).
In the superbubble, the SN ejecta can freely move away 
at a very high speed,
and the blast shock will propagate with a small Mach number
so that it can be expected to be weak and become thermalized,
with most of the SN kinetic energy carried by this part of the 
material being deposited
in the thermal energy of the hot medium
\citep{2005ApJ...628..205T}.

The MC with which the SNR probably interacts may survive the
strong radiation of the OB stars in the superbubble.
The photodissociation timescale of the MC can be
comparable to, or even larger than, the age of the superbubble.
For simplicity, the dissociating far ultraviolet (FUV) radiation of
the OB stars (of number $N_{\rm OB}$), to which the MC is exposed,
is approximated as being from the bubble center,
and the molecular column density along the bubble radial
is approximated to the line-of-sight column density $\NHH$. 
The distance from the MC to the center, $R_{\rm MC}$, is
not less than the projected distance to the center
($\sim22.5'$ or $\sim40\du$\,pc).
Each dissociating photon is absorbed by a hydrogen molecule
with a photodissociation probability, $p_{\rm D}$
$\sim0.15$ (\citealt{2011piim.book.....D}).
%meanwhile, each neutral atomic hydrogen yielded from the dissociation
%would absorb
%an FUV photon before the cloud is dissociated.
%That is, every incident FUV photon dissociates $p_{\rm D}/(1+2p_{\rm D})$
%H$_2$ molecules on average.
Thus, the photodissociation timescale of the MC is given by
%$\tau_{\rm MC}\ga (1+2p_{\rm D})\NHH(4\pi R_{\rm MC}^2)/
$\tau_{\rm MC}\ga \NHH(4\pi R_{\rm MC}^2)/
(p_{\rm D}N_{\rm OB}S_{\rm D})$,
where $S_{\rm D}$ is the mean production rate of Lyman band
dissociating photons of an OB star 
($\sim10^{47.5}\ps$, \citealt{1998ApJ...501..192D}).
It is estimated to be
$\tau_{\rm MC}\ga
6.4\E{6}(\NHH/1\E{21}\cm^{-2})
(N_{\rm OB}/20)^{-1}
(S_{\rm D}/10^{47.5}\ps)^{-1}
\du^2 \yr$,
where the reference value $1\E{21}\cm^{-2}$ is adopted for $\NHH$
according to 
Table~\ref{parameter} and Table~\ref{parameter2},
and $N_{\rm OB}\sim20$ is adopted as a reference value
considering this approximate number of stars are possibly inside
and responsible for the superbubble (based on Figure~\ref{obstar}).
On the other hand, the age of the superbubble is estimated to be
\citep{1977ApJ...218..377W}
$t_{\rm SB}\sim 5.8\E6 d_{6.1}^{5/3} (N_{\rm OB}/20)^{-1/3}
\times (L_w/10^{34} \erg\ps)^{-1/3} (n_{0}/0.6 
{\rm cm}^{-3})^{1/3} {\rm yr}$,
where $L_w$ is the power of the wind of an OB star,
and $n_0$ is the atomic number density of the environment medium
into which the superbubble expands.
Here we use a density of $0.6\cm^{-3}$ typical for a warm interstellar
HI gas (which takes the largest volume fraction next to the coronal gas in
the Galactic disk) to be a reference value for $n_0$
\citep{2011piim.book.....D}.
Thus their ratio is 
$\tau_{\rm MC}/t_{\rm SB}\ga
1.1\times (N({\rm H}_2)/1\E{21}{\rm cm}^{-2})
(N_{\rm OB}/20)^{-2/3}d_{6.1}^{1/3} 
(S_{\rm D}/10^{47.5} {\rm s}^{-1})^{-1}
(L_w/10^{34}\erg \ps)^{1/3} (n_0/0.6 {\rm cm}^{-3})^{-1/3} $.
Here we have ignored the formation of new molecules and the extinction
of the dissociating photons by dust grains in the molecular gas.
%The FUV optical depth caused by the dust opacity in the outer region
%of the molecular clouds can be of unity
%\citep{1995ApJ...455..133H},
%which may triple the estimate of the photodissociation timescale. 
This estimate indicates that, 
with some typical or proper values of parameters adopted, 
the photodissociation timescale appears not less than
the age of the superbubble,
and thus the survival of the MC in the superbubble could be reasonable.

It has been suggested that \snr\ is inside another superbubble
found in HI emission at $\VLSR\sim -70 \km\ps$,
with a radius in the range $\sim38'$--$70'$,
centered at approximately R.A.$=20^{\rm h}16^{\rm m}15^{\rm s}$, 
decl.=$36^\circ 40'$ (J2000) \citep{1997A&A...317..212W}.
This superbubble, however, is not as favored as the $\sim-64\km\ps$ one
in view of the bigger offset of the LSR velocity $\sim-70\km\ps$
from the CO-line velocity of the associated MC ($\sim-58\km\ps$) than
that of $\sim-64\km\ps$.
Incidentally, we do not find the counterpart of this HI superbubble
in the \emph{WISE} near- and mid-IR observations.

\subsection{Radio and X-ray Configuration}\label{configuration}

The X-ray observation of \snr\ shows 
a cometary-like trailing structure, 
$\sim 200''\times 300''$
or $\sim 5.9\du\parsec\times 8.9\du\parsec$ in size
at 0.3--7\,keV \citep{2013ApJ...774...33M}.
The radio emission takes a blow-out, arc-like shape, with the X-ray trail at its
symmetric axis,
and has a remarkably larger size
($\sim6'\times8'$ or $10.6\du\parsec\times14.2 \du\parsec$, 
even up to $16'$)
than the X-ray emission and a different brightness peak location
from the X-ray one \citep{2003ApJ...588..852K,2013ApJ...774...33M}.
The apparent one-sided confinement of the X-ray emission represents
a typical structure for a relative oriented motion
between the ambient gas and the PWN.
But why is the radio emission much more extended than the X-ray nebula,
with different locations for the brightness peaks in the two bands?

\cite{2013ApJ...774...33M} suggest that the radio emission of \snr\
is a relic PWN that was crushed by the reverse shock propagating back
from somewhere external, and the X-ray nebula is the new trailing PWN
after the passage of the reverse shock and the subsequent oscillation
of the nebula.
The pulsar has moved for about 10 kyr southeastward
from the explosion site, assumed to be at 
the location of the radio brightness peak.
However, as the authors point out, no sign of a forward shock
(especially in the nearby southeastern region)
has been found in this scenario.
Also, according to the simulation in \cite{2001ApJ...563..806B},
\cite{2004A&A...420..937V}, and
\cite{kolb17}, relic radio PWN are apparently swept/left behind 
the head of the new, small X-ray nebula.
In particular, given the  physical contact of the 
eastern and southwestern edges
of the extended radio emission with the MC, this seems to leave no room for
the reverse shock that shocks against and crushes the 
suggested radio relic PWN.

With the probable interaction of \snr\ with the ambient 
MC revealed here, 
we suggest instead that the radio emission 
is mainly a remnant of the blastwave
that propagates into the MC at $\VLSR\sim-58\km\ps$,
although it may include a contribution from the PWN.
This scenario also allows for a reverse shock to have moved backward
from the remnant's edge and crushed the PWN, forming the trailing
morphology of X-ray emission, as in the case of other, similar PWNe
(e.g.\ N157B, \citealt{2004A&A...420..937V,chenyang06}).
This scenario naturally addresses the question, as noted in
\cite{2013ApJ...774...33M},
of the lack of the X-ray emission related to the ejecta and blastwave.

The progenitor exploded somewhere along the symmetric axis of the radio
emission or the X-ray trail, with a part of the blastwave shocking against the MC and
the other part expanding, and quickly becoming sufficiently faint 
in a low-density region, possibly a portion of the superbubble.
The shock in the MC decelerated and entered the intense radiative
stage shortly thereafter, with the shocked gas 
at that point at too low a temperature to
emit X-rays.
Actually, the cooling time scale of the shock propagating into the
molecular gas is 
$ t_{\rm cool}=2.8\E{3} (E/10^{51}\erg)^{0.24} 
(n_{\rm a}/80\cm^{-3})^{-0.52}
$\,yr
\citep{1981MNRAS.195.1011F},
where $E$ is the SNR's explosion energy and $n_{\rm a}$ is the preshock H atom density.
For an average molecular density $n({\rm H}_2)\sim40\cm^{-3}$
(see Tables~\ref{parameter} and \ref{parameter2}),
the cooling time is much shorter than the spin down time $\sim1\E{4}\yr$
\citep{2013ApJ...774...33M}.
Although the SNR may have a part blown out, for simplicity
we crudely estimate the evolution of the radiative part
in the MC as a complete sphere with the radius as 
$r_s\sim5(\epsilon/0.24)^{5/21}(E/10^{51}\erg)^{5/21}
(n_{\rm a}/80\cm^{-3})^{-5/21}(t/10^{4}\yr)^{2/7}\parsec,
$
where $\epsilon\sim0.24$ is the energy fraction
\citep{mckee77,blinnikov82}.
If we approximate the pulsar's spindown time of $\sim10^{4}\yr$ as the age
of the remnant and adopt an explosion energy $\sim10^{51}\erg$,
then the radius is $\sim5\parsec$,
very similar to the size of the radio emission.
%The shock velocity
%is $\sim(2/7)r_s/t\sim140(r_s/5\parsec)(t/10^4\yr)^{-1}\km\ps$.
The shock velocity is 
$ v_s \sim (2/7)r_s/t \sim140(\epsilon/0.24)^{5/21}
(E/10^{51}\erg)^{5/21}(n_{\rm a}/80\cm^{-3})^{-5/21}(t/10^{4}\yr)^{-5/7}
\km\ps.
$
Since MCs are usually highly clumpy, 
the observed shocked molecular gas showing broad CO-line wings should be
in dense clumps, in which the shock velocity can be well below $50\km\ps$
\citep{draine93} so that the molecules are not dissociated.
The radio emission as a relic of the blastwave in this scenario indicates
a composite nature for SNR~\snr.

\section{Summary} \label{summarize}
We have performed a new millimeter CO-line observation toward the
region of \snr, which was thought to be a filled-center type SNR,
and an optical investigation of the coincident superbubble.
The CO observation shows that
the SNR delineated by the radio emission is projectively
covered by a bar-like molecular structure
at $\VLSR=-56.5$ to $-55.5\km\ps$.
Both the symmetric axis of the radio emission and the trailing
X-ray PWN appear projectively to be at a gap
between two molecular gas patches at $-58$ to $-57\km\ps$.
Asymmetric broad line profiles of the $\sim-58\km\ps$ \twCO-line
are obtained from the
molecular gas at the eastern and southwestern boundary of
the radio emission.
This could well be a kinematic evidence of the physical contact
between \snr\ and the $\sim-58\km\ps$ ($-60$ to $-54\km\ps$) MC complex.
A superbubble, $\sim37'$ in angular radius and centered at
($20^{\rm h}14^{\rm m}03^{\rm s}$, $37^{\circ}13'27''$, J2000),
seemingly surrounding the SNR, is found in 
HI 21cm ($\VLSR\sim-61$ to $-68\km\ps$), \emph{WISE} mid-IR,
and optical extinction observations.
We built a multi-band photometric stellar sample of over 
0.18~million
stars within the superbubble region and found 82 OB star candidates.
The distribution of the stars' distances is 
likely peaked at 7.3\,kpc
(with a dispersion of 1.9\,kpc)
%(within a 20\% error) 
and seems consistent with
the previously suggested distance of $6.1\pm0.9\kpc$ for \snr.
We suggest the arc-like radio emission is mainly a relic of
the part of the blastwave
that is driven into the MC complex, and is now in a radiative stage,
while the other part of the blastwave has been expanding into
the low-density region, very likely in the superbubble.
This scenario naturally explains the lack
of X-ray emission related to the ejecta and blastwave.
The SNR-MC interaction also favors a hadronic contribution
to the $\gamma$-ray emission from the \snr\ region.

\begin{acknowledgements}
We are grateful to the staff of Qinghai Radio Observing Station
at Delingha for their help during the observation.
We highly appreciate Xin Zhou and Gao-Yuan Zhang
for the advice on the data analysis, and Xiao Zhang for the
discussion about PWN physics.
This work is supported by
the 973 Program grants 2015CB857100 and 2017YFA0402600
and NSFC grants 11233001, 11633007, 11773014, 11503008, and 11590781.
This research has made use of the NVSS data and
NASA's Astrophysics Data System\footnote{http://adswww.harvard.edu/}.

\end{acknowledgements}

%\bibliography{ref.bib}
%\bibliographystyle{apj}

\newpage

\begin{deluxetable}{ccccc}
\tabletypesize{\scriptsize}
%\rotate
\tablecaption{Fitted and Derived Parameters for
the MCs around $-58\km\ps$ in Region ``R''\tablenotemark{a}
\label{parameter}}
\tablewidth{0pt}
\tablehead{
\multicolumn{5}{c}{Gaussian Components} \\
\cline{1-5} \\
\colhead{Line} & \colhead{Center ($\km\ps$)} & \colhead{FWHM ($\km\ps$)} & 
\colhead{$T_{\rm peak}$(K)} & \colhead{$W$(K\,$\km\ps$)}  \\
}
\startdata
\twCO (\otz) & $-57.5$ & 2.6 & 3.6 & 9.9 \\
\thCO (\otz) & $-57.1$ & 2.0 & 0.6 & 1.3  \\
\cutinhead{Molecular Gas Parameters}
$N$(H$_2$)($10^{21}$cm$^{-2}$)\tablenotemark{b} & $n(\mbox{H}_2)\du(\cm^{-3})$ 
& $M\du^{-2}(10^3M_\odot)$\tablenotemark{b} 
& $T_{\rm ex}$(K)\tablenotemark{c} & $\tau$(\thCO)\\
\hline \\
1.4/1.8 & 34/43 &  4.1/5.1 & 9.5 & 0.18 
\enddata
\tablecomments{}
\tablenotetext{a}{The region is defined in Figure~\ref{overall}.}
\tablenotetext{b}{See the text for the two estimating methods.}
\tablenotetext{c}{The excitation temperature is calculated from
the maximum \twCO (\otz) emission point of the $-58 \km\ps$ component.}
\end{deluxetable}

\begin{deluxetable}{ccccc}
\tabletypesize{\scriptsize}
\tablecaption{Fitted and Derived Parameters for
the MCs around $-58\km\ps$ in Region ``L''\tablenotemark{a}
\label{parameter2}}
\tablewidth{0pt}
\tablehead{
\multicolumn{5}{c}{Gaussian Components} \\
\cline{1-5} \\
\colhead{Line} & \colhead{Center ($\km\ps$)} & \colhead{FWHM ($\km\ps$)} & 
\colhead{$T_{\rm peak}$(K)} & \colhead{$W$(K\,$\km\ps$)}  \\
}
\startdata
\twCO (\otz) & $-57.5$ & 2.1 & 2.9 & 6.4 \\
\thCO (\otz) & $-57.4$ & 0.8 & 0.6 & 0.5  \\
\cutinhead{Molecular Gas Parameters}
$N$(H$_2$)($10^{21}$cm$^{-2}$)\tablenotemark{b} & $n(\mbox{H}_2)\du(\cm^{-3})$ 
& $M\du^{-2}(10^3M_\odot)$\tablenotemark{b} 
& $T_{\rm ex}$(K)\tablenotemark{c} & $\tau$(\thCO)\\
\hline \\
0.4/1.1 & 21/57 &  0.3/0.8 & 7.5 & 0.21 
\enddata
\tablecomments{}
\tablenotetext{a,b,c}{Same notations as in Table~\ref{parameter}}
\end{deluxetable}

\clearpage

\label{lastpage}
\end{document}